\documentclass[aps,preprint]{revtex4}%
\usepackage{amsfonts}
\usepackage{amsmath}
\usepackage{amssymb}
\usepackage{subfigure}
\usepackage{graphicx}%
\begin{document}
\preprint{CTP-SCU/2017038}
\title{Holographic DC Conductivity for a Power-law Maxwell Field}
\author{Benrong Mu$^{a,b}$}
\email{benrongmu@cdutcm.edu.cn}
\author{Peng Wang$^{b}$}
\email{pengw@scu.edu.cn}
\author{Haitang Yang$^{b}$}
\email{hyanga@scu.edu.cn}
\affiliation{$^{a}$Physics Teaching and Research Section, College of Medical Technology,
Chengdu University of Traditional Chinese Medicine, Chengdu 611137, China }
\affiliation{$^{b}$Center for Theoretical Physics, College of Physical Science and
Technology, Sichuan University, Chengdu, 610064, PR China}

\begin{abstract}
We consider a neutral and static black brane background with a probe
power-law Maxwell field. Via the membrane paradigm, an expression for the
holographic DC conductivity of the dual conserved current is obtained. We also
discuss the dependence of the DC conductivity on the temperature, charge
density and spatial components of the external field strength in the boundary
theory. Our results show that there might be more than one phase in the
boundary theory. Phase transitions could occur where the DC conductivity or
its derivatives are not continuous. Specifically, we find that one phase
possesses a charge-conjugation symmetric contribution, negative
magneto-resistance and Mott-like behavior.

\end{abstract}
\maketitle
\tableofcontents

\bigskip



\section{Introduction}

The idea of the membrane paradigm was started by Thibault Damour
\cite{IN-Damour:1978cg}\ and then developed further by Kip Thorne et al.
\cite{IN-Thorne:1986iy,IN-Price:1986yy}. Later, a more systematic action-based
derivation was proposed by Parikh and Wilczek in \cite{IN-Parikh:1997ma},
which could apply to various field theories. In the membrane paradigm, the
observer at infinity sees that the black hole is equivalent with a thin fluid
membrane living just outside the black hole's event horizon, and hence the
black hole can be replaced by the fluid membrane. The membrane paradigm was
originally proposed to study astrophysical black holes
\cite{IN-Masso:1998fi,IN-Komissarov:2004ms,IN-Penna:2013rga}. Realizing the
membrane fluid could provide the long wavelength description of the strongly
coupled quantum field theory at a finite temperature, researchers take a new
interest in the membrane paradigm in the context of gauge/gravity duality
\cite{IN-Kovtun:2003wp,IN-Kovtun:2004de,IN-Iqbal:2008by,IN-Bredberg:2010ky}.
In \cite{IN-Iqbal:2008by}, the low frequency limit of the boundary theory
transport coefficients could be expressed in terms of geometric quantities
evaluated at the horizon by identifying the currents in the boundary theory
with radially independent quantities in bulk. \ The method of
\cite{IN-Iqbal:2008by} was later extended to calculate the DC conductivity in
the presence of momentum dissipation
\cite{IN-Blake:2013bqa,IN-Donos:2014cya,IN-Cremonini:2016avj,IN-Bhatnagar:2017twr}%
, where the zero mode of the current, not the current itself, did not evolve
in the radial direction. Specifically, the DC thermoelectric conductivity has
recently been obtained by solving a system of Stokes equations on the black
hole horizon for a charged fluid in Einstein-Maxwell theory
\cite{IN-Donos:2015gia}.

Nonlinear electrodynamics (NLED) is an effective model incorporating quantum
corrections to Maxwell electromagnetic theory. NLED is interesting per se, for
example some models give finite self-energy of charged particles and can
remove singularity at the classical level. Two famous NLED are
Heisenberg--Euler effective Lagrangian \cite{IN-Heisenberg:1935qt} and
Born-Infeld electrodynamics \cite{IN-Born:1934gh}. On the other hand, it is
well-known that the Maxwell action enjoys the conformal invariance in four
dimensions. A natural extension of the Maxwell action in $\left(  d+1\right)
$-dimensional spacetime that is the conformally invariant is the action of a
power-law Maxwell field \cite{IN-Hassaine:2007py}:%
\begin{equation}
S=\int d^{d+1}x\sqrt{-g}s^{p}\equiv\int d^{d+1}x\sqrt{-g}\mathcal{L}\left(
s\right)  \text{,} \label{eq:actionPL}%
\end{equation}
where we define a nontrivial scalar%
\begin{equation}
s=-\frac{1}{4}F^{ab}F_{ab}\text{;}%
\end{equation}
$F_{ab}=\partial_{a}A_{b}-\partial_{b}A_{a}$ is the electromagnetic field
tensor, and $A_{a}$ is the electromagnetic potential. The action $\left(
\ref{eq:actionPL}\right)  $ is conformally invariant provided $p=\left(
d+1\right)  /4.$ When $d=3$, the action $\left(  \ref{eq:actionPL}\right)  $
recovers the standard Maxwell action. However, we don't confine ourselves to
$p=\left(  d+1\right)  /4$ in our paper. Instead, we shall consider a more
general case from now on, in which $p$ is an arbitrary positive integer.
Coupling the power-law Maxwell field to gravity, various charged black holes
were derived in a number of papers
\cite{IN-Hassaine:2007py,IN-Maeda:2008ha,IN-Sheykhi:2012zz,In-Miskovic:2010ui,IN-Zangeneh:2015gja,IN-Hendi:2017mgb}%
. In the framework of gauge/gravity duality, holographic superconductors
\cite{IN-Jing:2011vz,IN-Jing:2015nqv}, action/complexity conjecture
\cite{IN-Wang:2017uiw}, and the DC conductivity in the massive gravity
\cite{IN-Dehyadegari:2017fqo} were studied in presence of a power-law Maxwell field.

This paper is a follow-up paper of our previous paper \cite{IN-Wang2017}. In
\cite{IN-Wang2017}, we used the method of \cite{IN-Iqbal:2008by} to compute
the DC conductivities of an conserved current dual to a probe nonlinear
electrodynamics field in a general neutral and static black brane background.
However, our previous paper dealt, primarily, with a NLED Lagrangian that
would reduce to the Maxwell-Chern-Simons Lagrangian for small fields. Clearly,
the power-law Maxwell field with $p\neq1$ does not belong to this class of
NLED models and would have some different predictions for the DC
conductivities in the boundary theory. For example, when the charge density
and magnetic field in the boundary theory vanish, the DC conductivities are
zero for $p\neq1$ in this paper while they are not in \cite{IN-Wang2017}.

In this paper, we will consider a neutral and static black brane background
with a probe power-law Maxwell field and the dual theory. The aim of this
paper is to find an expression for the holographic DC conductivity of the dual
conserved current and investigate the properties of the boundary theory, e.g
the\ possible phases and the magnetotransport. Note that the properties of
magnetotransport in holographic Dirac-Born-Infeld models have been discussed
in a probe case \cite{IN-Kiritsis:2016cpm} and taking into account the effects
of backreaction on the geometry \cite{In-Cremonini:2017qwq}.

The remainder of our paper is organized as follows: In section \ref{Sec:MP},
we briefly review the membrane paradigm for a power-law Maxwell field. The
holographic DC conductivity of the dual conserved current is studied in
section \ref{Sec:DCFGGD}. In section \ref{Sec:Con}, we conclude with a brief
discussion of our results. We use convention that the Minkowski metric has
signature of the metric $\left(  -+++\right)  \,$in this paper.

\section{Membrane Paradigm}

\label{Sec:MP}

In \cite{IN-Wang2017}, the electromagnetic membrane properties have been
examined for a general NLED model via the method of \cite{IN-Parikh:1997ma}.
In this section, we first give a quick review of the membrane paradigm in the
framework of a power-law Maxwell field. In membrane paradigm, a time-like
hypersurface, namely the stretched horizon, is put just outside the black hole
horizon. The stretched horizon is composed of a family of fiducial observers
with world lines $U^{a}$ and possesses a spacelike outward pointing normal
vector $n_{a}$. The stretched horizon is denoted by $\mathcal{S}$. To derive
the Euler-Lagrange equations from the action restricted to the spacetime
outside the stretched horizon $S_{\text{out}}$, it is necessary to add a
surface term $S_{\text{surf}}$ to $S_{\text{out}}$ to exactly cancel all the
boundary terms. Consequently, the total action can be rewritten as
\begin{equation}
S_{\text{tot}}=\left(  S_{\text{out}}+S_{\text{surf}}\right)  +\left(
S_{\text{in}}-S_{\text{surf}}\right)  ,
\end{equation}
where $\delta S_{\text{out}}+\delta S_{\text{surf}}=0$ will give the correct
equations of motion outside $\mathcal{S}$.

For a power-law Maxwell field $A_{a}$, the external action in a $\left(
d+1\right)  $-dimensional spacetime is given by the action $\left(
\ref{eq:actionPL}\right)  $. To cancel the boundary contribution on the
stretched horizon from the action $\left(  \ref{eq:actionPL}\right)  $, we add
a surface term $S_{\text{surf}}$%
\begin{equation}
S_{\text{surf}}=\int\limits_{\mathcal{S}}d^{3}x\sqrt{\left\vert h\right\vert
}j_{\text{s}}^{a}A_{a},
\end{equation}
where $h_{ab}=g_{ab}-n_{a}n_{b}$ is the induced metric on $\mathcal{S}$; we
define%
\begin{equation}
j_{\text{s}}^{a}=G^{ab}n_{b}, \label{eq:jas}%
\end{equation}
and
\begin{equation}
G^{ab}=-\frac{\partial\mathcal{L}\left(  s\right)  }{\partial F_{ab}}%
=ps^{p-1}F^{ab}. \label{eq:Gab}%
\end{equation}
Note that $j_{\text{s}}^{a}$ can be interpreted as the membrane current on the
stretched horizon since $n_{a}j_{\text{s}}^{a}=0$. This current corresponds to
the surface electric charge density $\rho=-j_{\text{s}}^{a}U_{a}$ and current
density $\mathbf{j}_{\text{s}}^{a}=j_{\text{s}}^{a}-\sigma U^{a}$.

In this paper, we consider a general black brane background, the metric of
which takes the form%
\begin{align}
ds^{2}  &  =g_{ab}dx^{a}dx^{b}=g_{rr}\left(  r\right)  dr^{2}+g_{\mu\nu
}\left(  r\right)  dx^{\mu}dx^{\nu}\nonumber\\
&  =-g_{tt}\left(  r\right)  dt^{2}+g_{rr}\left(  r\right)  dr^{2}%
+g_{zz}\left(  r\right)  \delta_{AB}dx^{A}dx^{B}, \label{eq:metricd}%
\end{align}
where indices $\left\{  a,b\right\}  $ run over the $\left(  d+1\right)
$-dimensional bulk space, $\left\{  \mu,\nu\right\}  $ over $d$-dimensional
constant-$r$ slice, and $\left\{  A,B\right\}  $ over spatial coordinates. We
assume that there is an event horizon at $r=r_{h}$, where $g_{tt}\left(
r\right)  $ has a first order zero, $g_{rr}\left(  r\right)  $ has a first
order pole, and $g_{zz}\left(  r\right)  $ is nonzero and finite. The Hawking
temperature of this black brane is
\begin{equation}
T=\frac{\sqrt{g_{tt}^{\prime}\left(  r_{h}\right)  g^{rr\prime}\left(
r_{h}\right)  }}{4\pi}. \label{eq:HT}%
\end{equation}
Now put the stretched horizon at $r=r_{0}$ with $r_{0}-r_{h}\ll r_{h}$. This
stretched horizon has%
\begin{equation}
n_{a}=\sqrt{g_{rr}\left(  r_{0}\right)  }\delta_{ar}\text{ and }U_{a}%
=-\sqrt{g_{tt}\left(  r_{0}\right)  }\delta_{at}.
\end{equation}
Thus, the membrane current $\left(  \ref{eq:jas}\right)  $ reduces to%
\begin{equation}
j_{\text{s}}^{\mu}=\sqrt{g_{rr}\left(  r_{0}\right)  }G^{\mu r}.
\label{eq:mur}%
\end{equation}
It showed in \cite{IN-Wang2017} that, on the stretched horizon, the NLED field
strength has
\begin{equation}
F^{rA}\left(  r_{0}\right)  =-\sqrt{\frac{g_{tt}\left(  r_{0}\right)  }%
{g_{rr}\left(  r_{0}\right)  }}F^{tA}\left(  r_{0}\right)  \text{.}
\label{eq:rt}%
\end{equation}
We then use eqns. $\left(  \ref{eq:Gab}\right)  $, $\left(  \ref{eq:mur}%
\right)  $ and $\left(  \ref{eq:rt}\right)  $ to rewrite $j_{\text{s}}^{A}$
as
\begin{equation}
j_{\text{s}}^{A}=ps^{p-1}\left(  r_{0}\right)  E^{A}, \label{eq:mcurrent}%
\end{equation}
where the electric field measured by the fiducial observers on the stretched
horizon is%
\begin{equation}
E^{a}=F^{ta}\left(  r_{0}\right)  \sqrt{g_{tt}\left(  r_{0}\right)  },
\end{equation}
and $s$ on the stretched horizon becomes%
\begin{equation}
s\left(  r_{0}\right)  =\frac{1}{2}\left[  E^{r}E_{r}-\frac{F^{AB}\left(
r_{0}\right)  F_{AB}\left(  r_{0}\right)  }{2}\right]  .
\end{equation}
From eqn. $\left(  \ref{eq:mcurrent}\right)  $, we can read that the diagonal
components of the conductivities of the stretched horizon are
\begin{equation}
\sigma_{s}^{AA}\equiv\sigma_{s}=ps^{p-1}\left(  r_{0}\right)  \text{,}%
\end{equation}
and the Hall components are zero. These fields also increase the black hole's
entropy $S$ in accord with the Joule-heating relation \cite{IN-Price:1986yy}:%
\begin{equation}
T\frac{dS}{dt}=\alpha^{2}\int\limits_{\mathcal{S}}dA\sum_{B}j_{\text{s}}%
^{B}E^{B}=\alpha^{2}\int\limits_{\mathcal{S}}dA\sigma_{s}\sum_{B}\left(
E^{B}\right)  ^{2}, \label{eq:dS/dt}%
\end{equation}
where $\alpha=\sqrt{g_{tt}\left(  r_{0}\right)  }$ is the renormalized factor
\cite{IN-Price:1986yy}.

\section{DC Conductivity From Gauge/Gravity Duality}

\label{Sec:DCFGGD}

We now consider a probe power-law Maxwell field in the background of a
$\left(  d+1\right)  $-dimensional black brane with the metric $\left(
\ref{eq:metricd}\right)  $. For simplicity, we assume that this black brane is
uncharged with trivial background configuration of the power-law Maxwell
field. This power-law Maxwell field is a U$\left(  1\right)  $ gauge field and
dual to a conserved current $\mathcal{J}^{\mu}$ in the boundary theory. The
corresponding AC conductivities are given by%
\begin{equation}
\left\langle \mathcal{J}^{A}\left(  k_{\mu}\right)  \right\rangle =\sigma
^{AB}\left(  k_{\mu}\right)  F_{Bt}\left(  r\rightarrow\infty\right)  ,
\end{equation}
where the boundary theory lives at $r\rightarrow\infty$. The DC conductivities
are obtained in the long wavelength and low frequency limit:
\begin{equation}
\sigma_{D}^{AB}=\lim_{\omega\rightarrow0}\lim_{\vec{k}\rightarrow0}\sigma
^{AB}\left(  k_{\mu}\right)  .
\end{equation}
We can compute the expectation value of the current $\mathcal{J}^{\mu}$ for
the boundary theory by \cite{IN-Wang2017}%
\begin{equation}
\left\langle \mathcal{J}^{\mu}\right\rangle =\Pi^{A}\equiv\frac{\partial
\mathcal{L}\left(  s\right)  }{\partial\left(  \partial_{r}A_{\mu}\right)
}|_{r\rightarrow\infty}=-\sqrt{-g}G^{r\mu}|_{r\rightarrow\infty},
\end{equation}
where $\Pi^{\mu}$ is the conjugate momentum of the field $A_{\mu}$ with
respect to $r$-foliation. When $\mu=t$, one hence has%
\begin{equation}
\rho=\left\langle \mathcal{J}^{t}\right\rangle =-\sqrt{-g}G^{rt}%
|_{r\rightarrow\infty},
\end{equation}
where $\rho$ can be interpreted as the charge density in the dual field theory.

Identifying the currents in the boundary theory with radially independent
quantities in the bulk, authors of \cite{IN-Iqbal:2008by} showed that the
membrane paradigm fluid on the stretched horizon determined the low frequency
limit of conductivities of a conserved current in the boundary theory, which
was dual to a Maxwell field in bulk. Later, the method of
\cite{IN-Iqbal:2008by} was extended to the NLED case in \cite{IN-Wang2017}. In
particular, it showed there that, in the long wavelength and low frequency
limit, i.e. $\omega\rightarrow0$ and $\vec{k}\rightarrow0$ with $F_{\rho
\sigma}$ and $\Pi^{\eta}$ fixed, the following quantities did not evolve in
the radial direction:
\begin{equation}
\partial_{r}\Pi^{\mu}=0\text{ and }\partial_{r}F_{\mu\nu}=0.
\end{equation}
On the horizon, we have%
\begin{equation}
\Pi^{A}\left(  r_{h}\right)  =\sqrt{-g}\frac{j_{\text{s}}^{A}}{\sqrt
{g_{rr}\left(  r_{h}\right)  }}=g_{zz}^{\frac{d-3}{2}}\left(  r_{h}\right)
\mathcal{L}^{\prime}\left(  s\right)  |_{r=r_{h}}F_{At},
\end{equation}
where we take the limit $r_{0}\rightarrow r_{h}$. Here, $s$ on the horizon
becomes%
\begin{equation}
s\left(  r_{h}\right)  =\frac{1}{2}\left[  \eta F^{rt}\left(  r_{h}\right)
^{2}-\frac{B^{2}}{g_{zz}^{2}\left(  r_{h}\right)  }\right]  , \label{eq:srh}%
\end{equation}
where we define%
\begin{equation}
\eta\equiv g_{rr}\left(  r_{h}\right)  g_{tt}\left(  r_{h}\right)  \text{ and
}B^{2}\equiv\frac{1}{2}\sum\limits_{A,B}F_{AB}^{2}.
\end{equation}
For $d=2$ and $3$, the magnetic field is a scalar and a vector, respectively,
and $B$ can be treated as the magnitude of the magnetic field in the boundary
theory. To express $F^{rt}\left(  r_{h}\right)  $ in terms of quantities in
the boundary theory, we can use the following formula%
\begin{equation}
\Pi^{t}\left(  r_{h}\right)  =\Pi^{t}\left(  r\rightarrow\infty\right)  =\rho.
\end{equation}
On the boundary, we have that, in the zero momentum limit,%
\begin{equation}
\left\langle \mathcal{J}^{A}\right\rangle =\Pi^{A}\left(  r\rightarrow
\infty\right)  =\Pi^{A}\left(  r_{h}\right)  =g_{zz}^{\frac{d-3}{2}}\left(
r_{h}\right)  \mathcal{L}^{\prime}\left(  s\right)  |_{r=r_{h}}F_{At}.
\label{eq:J}%
\end{equation}
From eqn. $\left(  \ref{eq:J}\right)  $, we can read that the diagonal
components of the DC conductivities in the boundary theory:
\begin{equation}
\sigma_{D}^{AA}\equiv\sigma_{D}\left(  \tilde{\rho},\tilde{B}\right)
=-\frac{g_{zz}^{\frac{d-3}{2}}\left(  r_{h}\right)  p}{2^{p-1}}\tilde
{B}^{2p-2}\frac{\tilde{\rho}/\tilde{B}^{2p-1}}{x\left(  \tilde{\rho}/\tilde
{B}^{2p-1}\right)  }\text{,} \label{eq:sigma}%
\end{equation}
where, for later convenience, we define%
\begin{equation}
\tilde{\rho}\equiv\frac{2^{p-1}\rho}{g_{zz}^{\frac{d-1}{2}}\left(
r_{h}\right)  p}\text{ and }\tilde{B}\equiv\frac{B}{g_{zz}\left(
r_{h}\right)  }\geq0\text{,}%
\end{equation}
and $x\left(  y\right)  $ is the inverse of the function $y\left(  x\right)
=-x\left(  x^{2}-1\right)  ^{p-1}$. Note that the Hall components vanish. The
value of $\sigma_{D}$ in the limit of $\tilde{B}\rightarrow0$ depends on the
value of $p$:%
\begin{equation}
\sigma_{D}=1\text{, for }p=1\text{, and }\sigma_{D}\left(  \tilde{\rho
},0\right)  =0\text{, otherwise. }%
\end{equation}
When $d=3$ and $p=1$, eqn. $\left(  \ref{eq:sigma}\right)  $ reproduces the
well-known result in the Maxwell case \cite{IN-Price:1986yy}%
\begin{equation}
\sigma_{D}=1\text{.}%
\end{equation}
For $p\neq1$, the DC conductivity $\sigma_{D}$ is zero in the absence of the
magnetic field and charge density in the boundary theory, which is consistent
with eqn. $\left(  44\right)  $ with $q=0$ in \cite{IN-Dehyadegari:2017fqo}.\

In the long wavelength and low frequency limit with $\omega\rightarrow0$ and
$\vec{k}\rightarrow0$, we keep $F_{\mu\nu}$ and $\Pi^{\mu}$ fixed and neglect
higher $\mu$-derivatives. This means that $F_{\mu\nu}$ and $\rho$ are constant
and homogeneous on the boundary. In this limit, one can relate the DC
conductivity $\sigma_{D}$ in the boundary theory to $\sigma_{s}$ of the
stretched horizon as%
\begin{equation}
\sigma_{s}=g_{zz}^{\frac{d-3}{2}}\left(  r_{h}\right)  \sigma_{D},
\end{equation}
which is also constant and homogeneous on the stretched horizon. Therefore in
the long wavelength and low frequency limit, the rate of the black hole's
entropy $S$ becomes%
\begin{equation}
T\frac{dS}{dt}=g_{zz}^{\frac{d-3}{2}}\left(  r_{h}\right)  \sigma_{D}%
\int\limits_{\mathcal{S}}\alpha^{2}\sum_{B}\left(  E^{B}\right)  ^{2}dA.
\end{equation}
The second law of black hole mechanics implies that the DC conductivity
$\sigma_{D}$ in the boundary theory is non-negative and real.

It is interesting to note that the function $x\left(  y\right)  $ is usually a
multivalued function, which indicates that there might exist more than one
phase and possible phase transitions. For later convenience, we define%
\begin{equation}
\tilde{\sigma}_{D}=\frac{2^{p-1}\sigma_{D}}{pg_{zz}^{\frac{d-3}{2}}\left(
r_{h}\right)  }.
\end{equation}

\subsection{Even Positive Integer $p$}

\begin{figure}[tb]
\begin{center}
\subfigure[{Plot of $y\left(  x\right)  =-x\left(  x^{2}-1\right)  $. On the green
segment, $x/y$ is positive and hence $\sigma_{D}$ becomes negative.}]{
\includegraphics[width=0.45\textwidth]{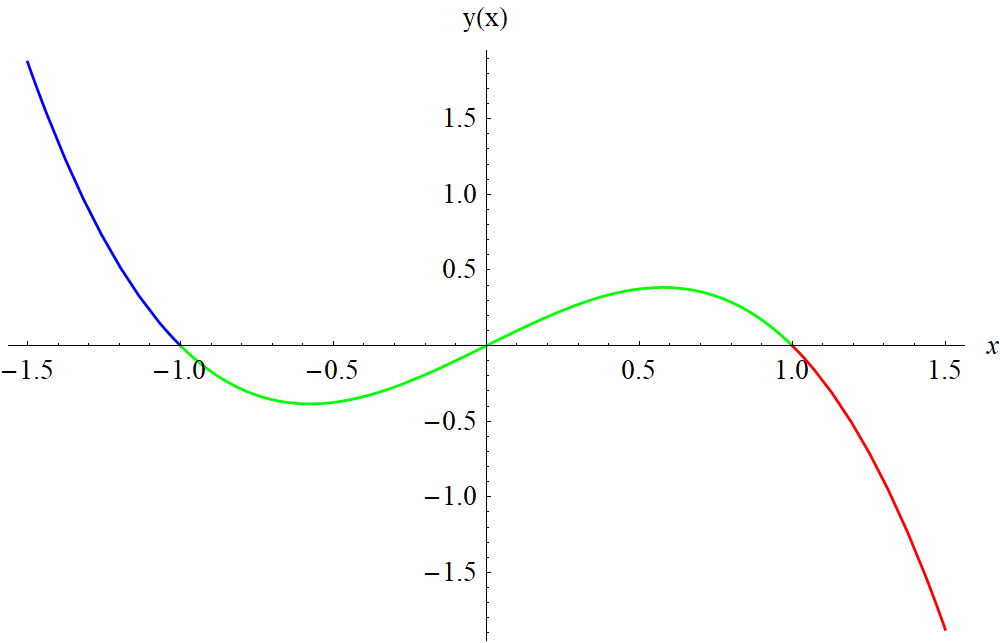}\label{fig:p2yvsx}}
\subfigure[{Plot of $x\left(  y\right)  $, the inverse function of $y\left(  x\right)  $.
Here we require that $\sigma_{D}$ is non-negative and real on $x\left(
y\right)  $.}]{
\includegraphics[width=0.45\textwidth]{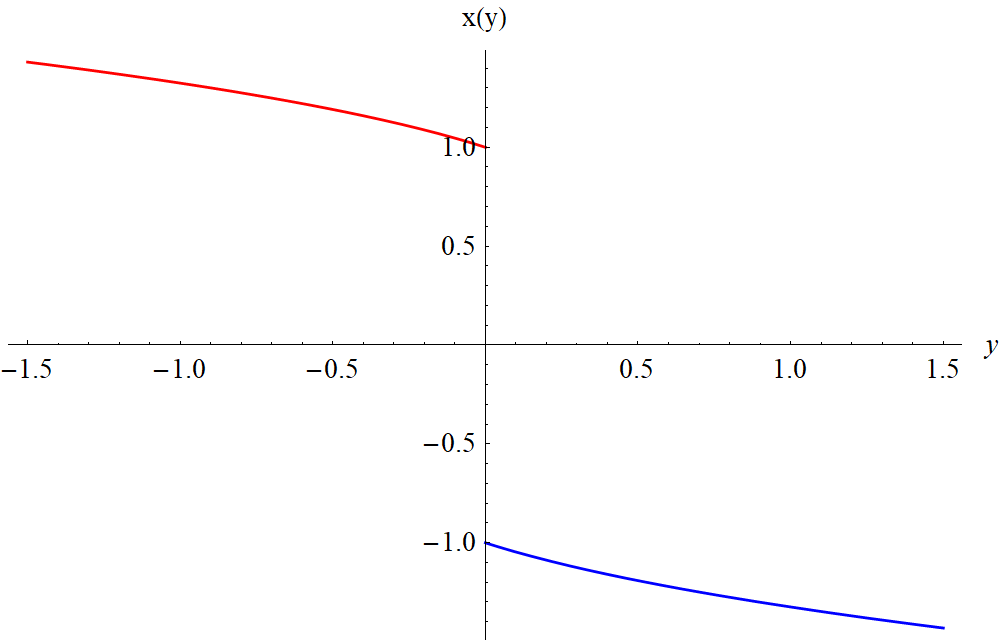}\label{fig:p2xvsy}}
\end{center}
\caption{Plots of $y\left(  x\right)  $ and $x\left(  y\right)  $ for $p=2$.}%
\end{figure}

\begin{figure}[tb]
\begin{center}
\includegraphics[width=0.5\textwidth]{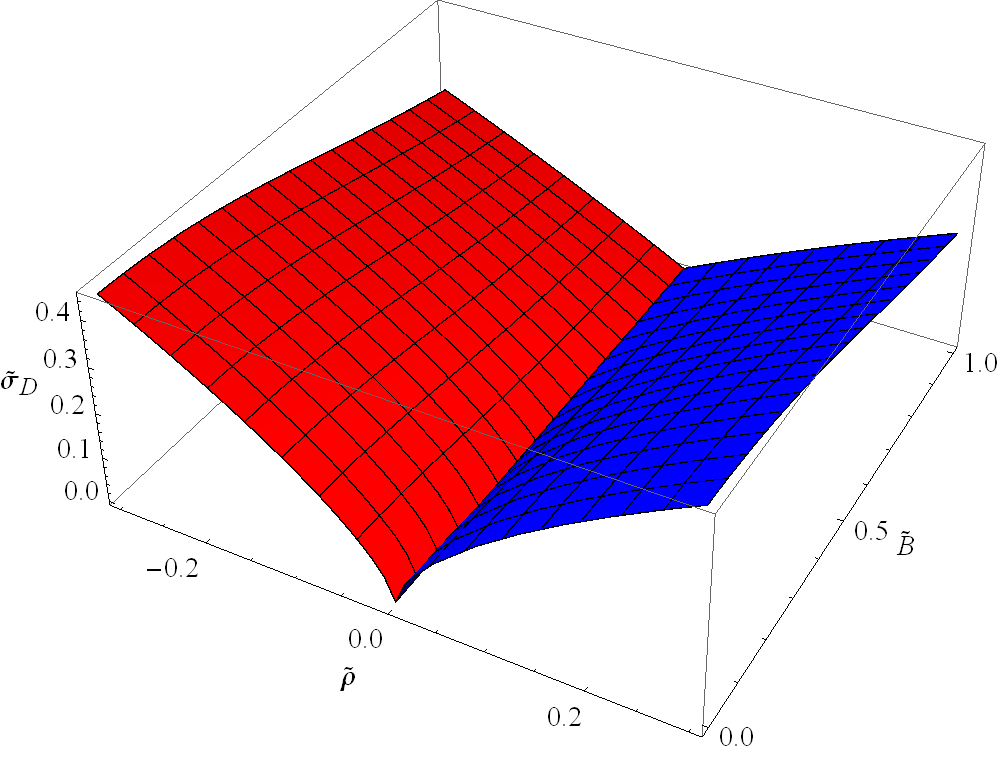}
\end{center}
\caption{Plot of $\tilde{\sigma}_{D}$ versus $\tilde{\rho}$ and $\tilde{B}$
for $p=2$. A continuous phase transition could occur at $\tilde{\rho}=0$,
where $\partial_{\tilde{\rho}}\tilde{\sigma}_{D}$ changes the sign.}%
\label{fig:DCp2}%
\end{figure}

We plot $y\left(  x\right)  =-x\left(  x^{2}-1\right)  ^{p-1}$ for $p=2$ in
FIG. \ref{fig:p2yvsx}. In fact, $y\left(  x\right)  $ and hence $\sigma_{D}$
in all the cases of $p$ being even positive integer show very similar behavior
as in that of $p=2$. So for concreteness, we shall focus on the case of $p=2$.
Bearing in mind that $\sigma_{D}$ is non-negative and real, eqn. $\left(
\ref{eq:sigma}\right)  $ shows that the green segment of $y\left(  x\right)  $
in FIG. \ref{fig:p2yvsx} is unphysical. Therefore, we only need to consider
the blue and red segments to find the inverse function of $y\left(  x\right)
$, which is plotted in FIG. \ref{fig:p2xvsy}. As shown in FIG.
\ref{fig:p2xvsy}, there is a discontinuity at $y=0$ for $x\left(  y\right)  $,
which, as will be shown later, indicates possible phase transitions at $y=0$.
Using $x\left(  y\right)  $ in FIG. \ref{fig:p2xvsy}, we plot $\tilde{\sigma
}_{D}$ versus $\tilde{\rho}$ and $\tilde{B}$ in FIG. \ref{fig:DCp2}. It shows
in FIG. \ref{fig:DCp2} that $\tilde{\sigma}_{D}$ is continuos everywhere but
the derivative $\partial_{\tilde{\rho}}\tilde{\sigma}_{D}$ changes the sign at
$\tilde{\rho}=0$. These observations imply that there might exist two phases
for $\tilde{\rho}>0$ and $\tilde{\rho}<0$, respectively, and a continuous
phase transition could occur at $\tilde{\rho}=0$.

Since $x\left(  0\right)  =1$, eqn. $\left(  \ref{eq:sigma}\right)  $ shows
that the DC conductivity $\sigma_{D}$ vanishes at zero charge density, which
implies that the main contribution to $\sigma_{D}$ is from momentum relaxation
for the charge carriers in the system. As shown in FIG. \ref{fig:DCp2},
$\sigma_{D}$ increases with increasing $\left\vert \rho\right\vert $ at
constant $B$, which is a feature similar to the Drude metal. For the Drude
metal, a larger charge density provides more available mobile charge carriers
to efficiently transport charge. At constant $\rho$, $\sigma_{D}$ decreases
with increasing $B$, which means a positive magneto-resistance.

\subsection{Odd Positive Integer $p$}

\begin{figure}[tb]
\begin{center}
\subfigure[{Plot of $y\left(  x\right)  =-x\left(  x^{2}-1\right)  ^{2}$, where each
colored segment has a single-valued inverse function. On $y\left(  x\right)
$, $-x/y$ and $\sigma_{D}$ are always non-negative.}]{
\includegraphics[width=0.45\textwidth]{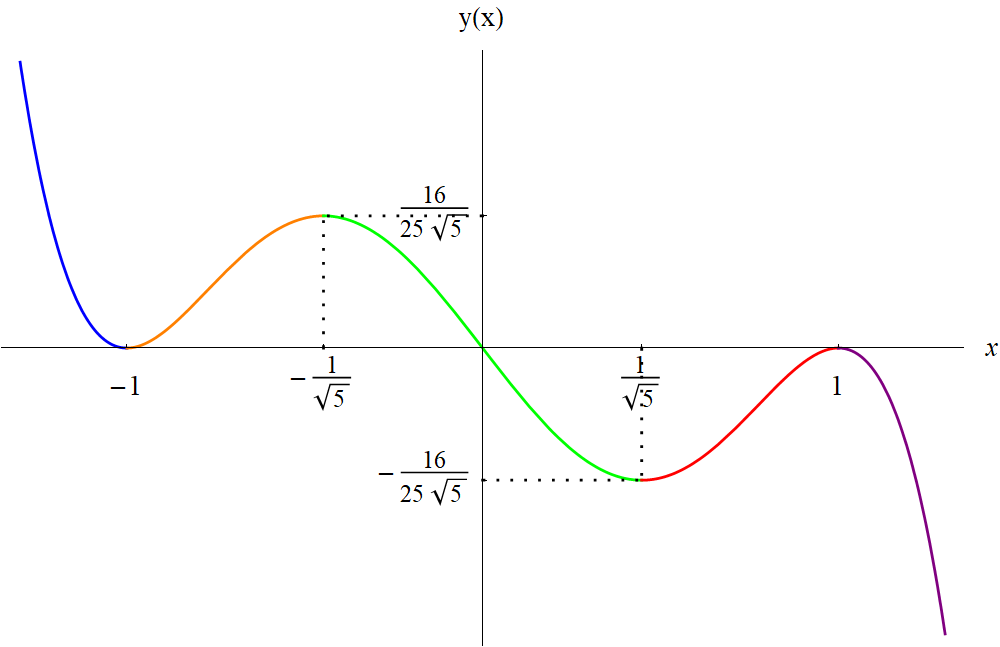}\label{fig:p3yvsx}}
\subfigure[{Plot of $x\left(  y\right)  $, the inverse function of $y\left(  x\right)  $.
Each colored single-valued segment corresponds to a possible phase in the
boundary theory.}]{
\includegraphics[width=0.45\textwidth]{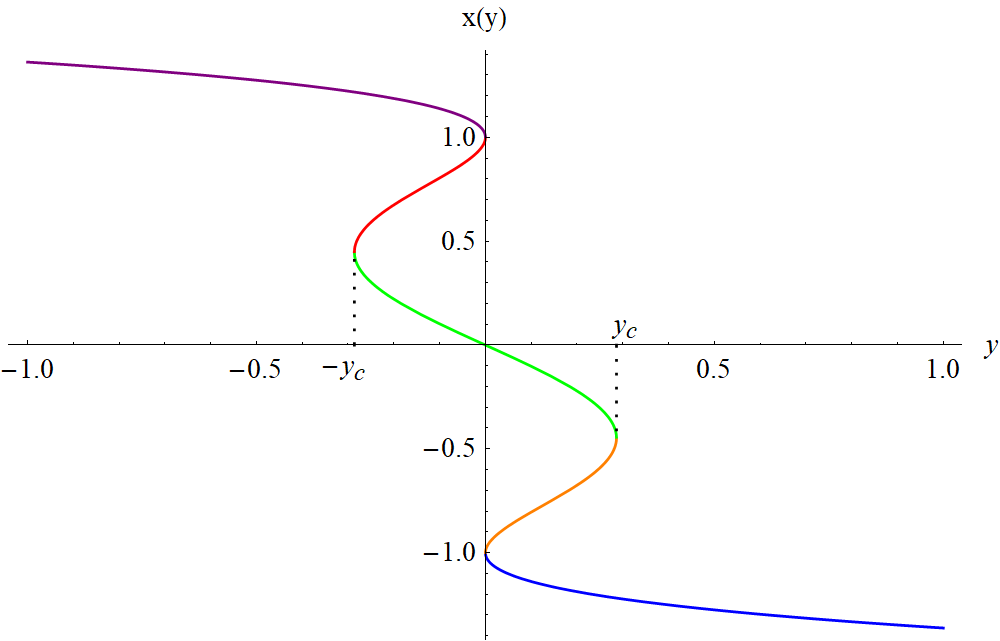}\label{fig:p3xvsy}}
\end{center}
\caption{Plots of $y\left(  x\right)  $ and $x\left(  y\right)  $ for $p=3$.}%
\label{fig:mr}%
\end{figure}

\begin{figure}[tb]
\begin{center}
\includegraphics[width=0.5\textwidth]{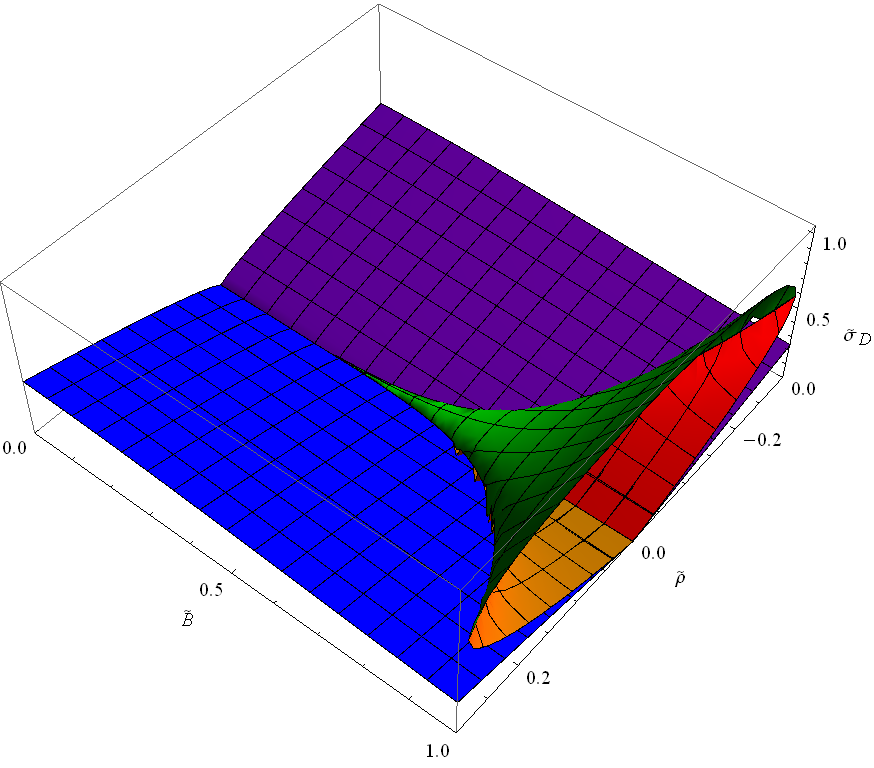}
\end{center}
\caption{Plot of $\tilde{\sigma}_{D}$ versus $\tilde{\rho}$ and $\tilde{B}$
for $p=3$. Five possible phases are represented by different colors. In the
region $y_{c}>\left\vert \tilde{\rho}/\tilde{B}^{5}\right\vert >0$, jumping
from one value of $\tilde{\sigma}_{D}$ to another can be considered as a first
order phase transition. Continuous phase transitions could occur at
$\tilde{\rho}=0$ and $\left\vert \tilde{\rho}/\tilde{B}^{5}\right\vert =y_{c}%
$.}%
\label{fig:DCp3}%
\end{figure}

Since all the cases with an odd positive integer $p$ share very similar
behavior, we shall focus on the case of $p=3$ here. The function $y\left(
x\right)  =-x\left(  x^{2}-1\right)  ^{p-1}$ for $p=3$ is shown in FIG.
\ref{fig:p3yvsx}. Unlike the $p=2$ case, $x/y\leq0$ and hence $\tilde{\sigma
}_{D}$ is non-negative for all points on $y\left(  x\right)  $ in the $p=3$
case. So the physical inverse function of $y\left(  x\right)  $ is plotted in
FIG. \ref{fig:p3xvsy}. As shown in FIG. \ref{fig:p3xvsy}, the function
$x\left(  y\right)  $ has a single value for $y^{2}>y_{c}^{2}$ \ and three
values for $y^{2}\leq y_{c}$. Here we define $y_{c}=\frac{16}{25\sqrt{5}}$. In
FIG. \ref{fig:p3xvsy}, $x\left(  y\right)  $ can be divided into $5$
single-valued segments: blue, orange, green, red, and purple ones, and each
segment corresponds to a possible phase in the boundary theory. In fact, we
have five possible phases: the blue phase exists for $\tilde{\rho}/\tilde
{B}^{5}\geq0$; the orange phase exists for $y_{c}\geq\tilde{\rho}/\tilde
{B}^{5}\geq0$; the green phase exists for $y_{c}\geq\tilde{\rho}/\tilde{B}%
^{5}\geq-y_{c}$; the red phase exists for $0\geq\tilde{\rho}/\tilde{B}^{5}%
\geq-y_{c}$; the purple phase exists for $\tilde{\rho}/\tilde{B}^{5}\leq0$. We
plot $\tilde{\sigma}_{D}$ versus $\tilde{\rho}$ and $\tilde{B}$ for the five
phases in FIG. \ref{fig:DCp3}. One can see that in the region $y_{c}%
>\left\vert \tilde{\rho}/\tilde{B}^{5}\right\vert >0$, three values of
$\tilde{\sigma}_{D}$ are allowed for fixed values of $\tilde{\rho}$ and
$\tilde{B}$. It means that $\tilde{\sigma}_{D}$ can jump from one value to
another. Since the value of $\tilde{\sigma}_{D}$ changes discontinuously, it
is acceptable to consider this transition as a first order phase transition.
On the other hand, the transitions occurring at $\tilde{\rho}=0$ and
$\left\vert \tilde{\rho}/\tilde{B}^{5}\right\vert =y_{c}$ can be regarded as
continuous phase transitions. To determine the stable phases and the
transition points, one needs to find the thermodynamic potential in a specific
boundary theory, which, however, is beyond the scope of this paper.

\begin{figure}[tb]
\begin{center}
\subfigure[{Plot of $\tilde{\sigma}_{D}$ versus $\tilde{B}$. The green lines, from left to
right, have $\tilde{\rho}=0$, $0.1$, $0.3$, $0.5$, and $0.7$. These lines show
that $\partial\tilde{\sigma}_{D}/\partial\tilde{B}>0$ and hence a negative magneto-resistance.}]{
\includegraphics[width=0.45\textwidth]{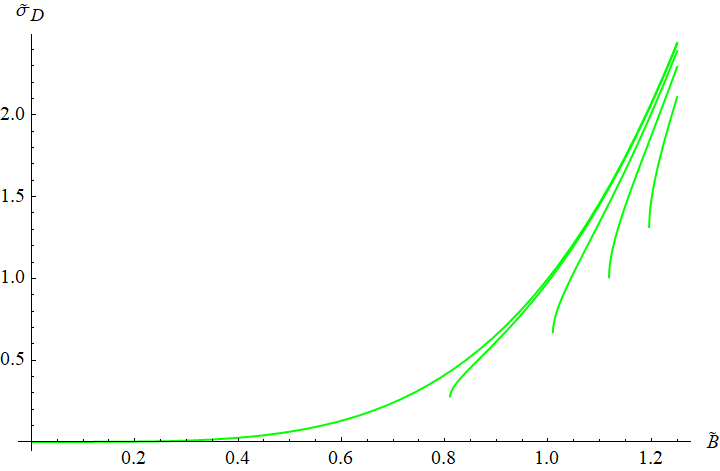}\label{fig:DCp3NM}}
\subfigure[{Plot of $\tilde{\sigma}_{D}$ versus $\tilde{\rho}$. The green lines, from
bottom to top, have $\tilde{B}=0.7$, $0.8$, $0.9$, $0.95$, and $1$. These
lines show that $\partial\tilde{\sigma}_{D}/\partial\rho<0$ and hence
Mott-like behavior.}]{
\includegraphics[width=0.45\textwidth]{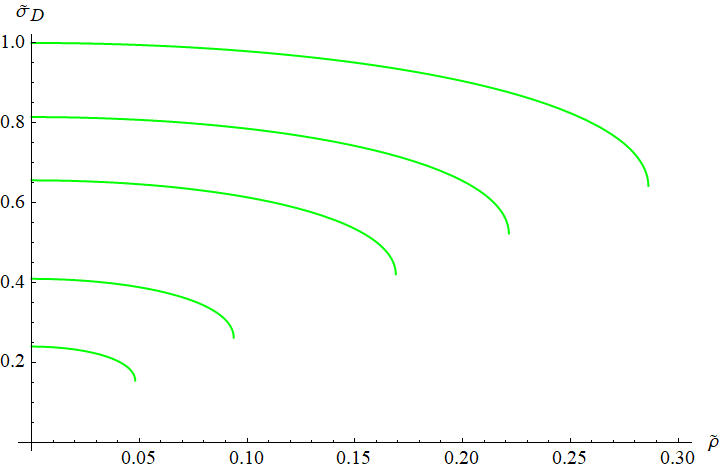}\label{fig:DCp3TJ}}
\end{center}
\caption{Plots of $\tilde{\sigma}_{D}$ versus $\tilde{B}$ and $\tilde{\rho}$,
respectively, for the green phase in the case of $p=3$.}%
\label{fig:DCp3NMTJ}%
\end{figure}

For the blue, orange, red, and purple phases, the behavior of the DC
conductivity $\sigma_{D}$ is similar to that in the $p=2$ case, i.e.
$\sigma_{D}=0$ at zero charge density, $\sigma_{D}$ increases with increasing
$\left\vert \rho\right\vert $ at constant $B$, and $\sigma_{D}$ decreases with
increasing $B$ at constant $\rho$. However, the green phase has some
interesting features:

\begin{itemize}
\item Charge conjugation symmetric contribution. At zero charge density,
$\sigma_{D}$ has a non-zero value, if $\tilde{B}\neq0$,
\begin{equation}
\sigma_{D}\left(  0,\tilde{B}\right)  =\frac{g_{zz}^{\frac{d-3}{2}}\left(
r_{h}\right)  p}{2^{p-1}}\tilde{B}^{2p-2},
\end{equation}
which can be interpreted by a incoherent contribution due to intrinsic current
relaxation and independent of the charge density. This contribution is also
known as the charge conjugation symmetric contribution
\cite{DCFGGD-Davison:2015bea,DCFGGD-Blake:2014yla}.

\item Negative magneto-resistance. We plot $\tilde{\sigma}_{D}$ versus
$\tilde{B}$ for $\tilde{\rho}=0$, $0.1$, $0.3$, $0.5$, and $0.7$ in FIG.
\ref{fig:DCp3NM}. FIGs. \ref{fig:DCp3} and \ref{fig:DCp3NM} show that
$\partial\tilde{\sigma}_{D}/\partial\tilde{B}>0$, which gives a negative
magneto-resistance at given temperature and charge density.

\item Mott-like behavior. We plot $\tilde{\sigma}_{D}$ versus $\tilde{\rho}$
for $\tilde{B}=0.7$, $0.8$, $0.9$, $0.95$, and $1$ in FIG. \ref{fig:DCp3TJ}.
Therefore we can see from FIGs. \ref{fig:DCp3} and \ref{fig:DCp3TJ} that
$\partial\tilde{\sigma}_{D}/\partial\rho<0$ for the green phase. This can be
explained by the electronic traffic jam: strong enough $e$-$e$ interactions
prevent the available mobile charge carriers to efficiently transport charges
\cite{DCFGGD-Baggioli:2016oju}. Note that a class of holographic models for
Mott insulators, whose gravity dual contained NLED, was studied in
\cite{DCFGGD-Baggioli:2016oju}.
\end{itemize}

\subsection{Temperature Dependence of DC Conductivity}

To discuss the temperature dependence of the DC conductivity, we can express
$\sigma_{D}$ in terms of $\rho$ and $B$:%
\begin{equation}
\sigma_{D}=-\frac{p\rho}{B}x^{-1}\left(  \frac{2^{p-1}\rho}{B^{2p-1}}%
g_{zz}^{\frac{4p-1-d}{2}}\left(  r_{h}\right)  \right)  . \label{eq:sigmaB}%
\end{equation}
When $d=4p-1$, the power-law Maxwell field action $\left(  \ref{eq:actionPL}%
\right)  $ is conformally invariant. In this case, the DC conductivity
$\sigma_{D}$ is independent of the geometric quantities evaluated at the
horizon, especially the Hawking temperature $T$ of the black brane. So the DC
conductivity $\sigma_{D}$ does not depend on the temperature of the boundary
theory when the power-law Maxwell field in bulk is conformally invariant. In
fact, the dual conserved current is also scale invariant. For this scale
invariant current at finite temperature, all nonzero temperatures should be
equivalent since there is no other scale with which to compare the temperature.

For $d\neq4p-1$, we can now discuss the temperature dependence of the DC
conductivity by relating $r_{h}$ to the Hawking temperature $T$. For
simplicity and concreteness, we consider the Schwarzschild AdS black brane%
\begin{equation}
ds^{2}=-\left(  r^{2}-r_{h}^{3}/r\right)  dt^{2}+\frac{dr^{2}}{\left(
r^{2}-r_{h}^{3}/r\right)  }+r^{2}\delta_{AB}dx^{A}dx^{B},
\end{equation}
where we take the AdS radius $L=1$, and $r_{h}$ determines the Hawking
temperature of the black hole:%
\begin{equation}
T=\frac{3r_{h}}{4\pi}\text{.}%
\end{equation}
Since $x\left(  y\right)  $ behaves differently for small $y$ depending on
whether $p$ is an even or odd integer, it will be convenient to consider these
two cases separately.

\begin{figure}[tb]
\begin{center}
\includegraphics[width=0.4\textwidth]{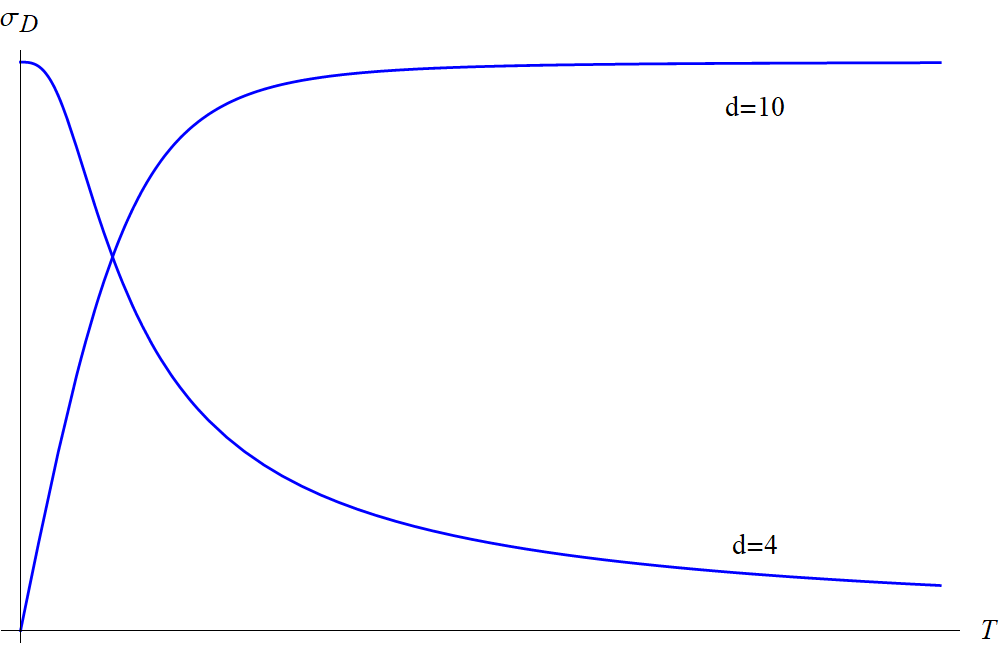}
\end{center}
\caption{Plot of $\sigma_{D}$ versus $T$ for $d=4$ and $10$ in the case of
$p=2$. Here the values of $\rho$ and $B$ are fixed with $\rho/B^{3}>0$, which
meas the bule phase.}%
\label{fig:DCT2}%
\end{figure}

When $p$ is even, one has that $x^{2}\left(  y\right)  \sim1$ for $y\ll1$. In
the case with $\rho=0$ and nonzero $B$, one has $\sigma_{D}=0$. In the case
with nonzero $B$ and $\rho$, one has that, when $d>4p-1$,%
\begin{equation}
\sigma_{D}\sim\rho^{\frac{2p-2}{2p-1}}T^{\frac{d+1-4p}{2p-1}}\text{ for small
}T\text{, and }\sigma_{D}\sim\frac{\left\vert \rho\right\vert }{B}\text{ for
large }T\text{,}%
\end{equation}
and when $d<4p-1$%
\begin{equation}
\sigma_{D}\sim\frac{\left\vert \rho\right\vert }{B}\text{ for small }T\text{,
and }\sigma_{D}\sim\rho^{\frac{2p-2}{2p-1}}T^{\frac{d+1-4p}{2p-1}}\text{ for
large }T\text{.}%
\end{equation}
FIG. \ref{fig:p2xvsy} shows that $x\left(  y\right)  $ is a monotonically
decreasing function of $y$. From eqn. $\left(  \ref{eq:sigmaB}\right)  $, we
find that $\partial\sigma_{D}/\partial T$ $<0$ for $d<4p-1$ and $\partial
\sigma_{D}/\partial T$ $>0$ for $d>4p-1$. If we define a metal and an
insulator for $\partial\sigma_{D}/\partial T<0$ and $\partial\sigma
_{D}/\partial T$ $>0$, respectively, one has a metal for $d<4p-1$ and an
insulator for $d>4p-1$. The results are summarized in TABLE \ref{tab:result}.
For fixed values of $\rho$ and $B$ with $\rho/B^{3}>0$, we plot $\sigma_{D}$
versus $T$ for $d=4$ and $10$ in the case of $p=2$ in FIG. \ref{fig:DCT2}.

\begin{figure}[tb]
\begin{center}
\subfigure[{Plot of $\sigma_{D}\left(  T\right)  $ in the the blue, orange and green
phases for $d=4$. For $T<T_{c}$, jumping from one value of $\sigma_{D}$ to
another represents a first order phase transition.}]{
\includegraphics[width=0.45\textwidth]{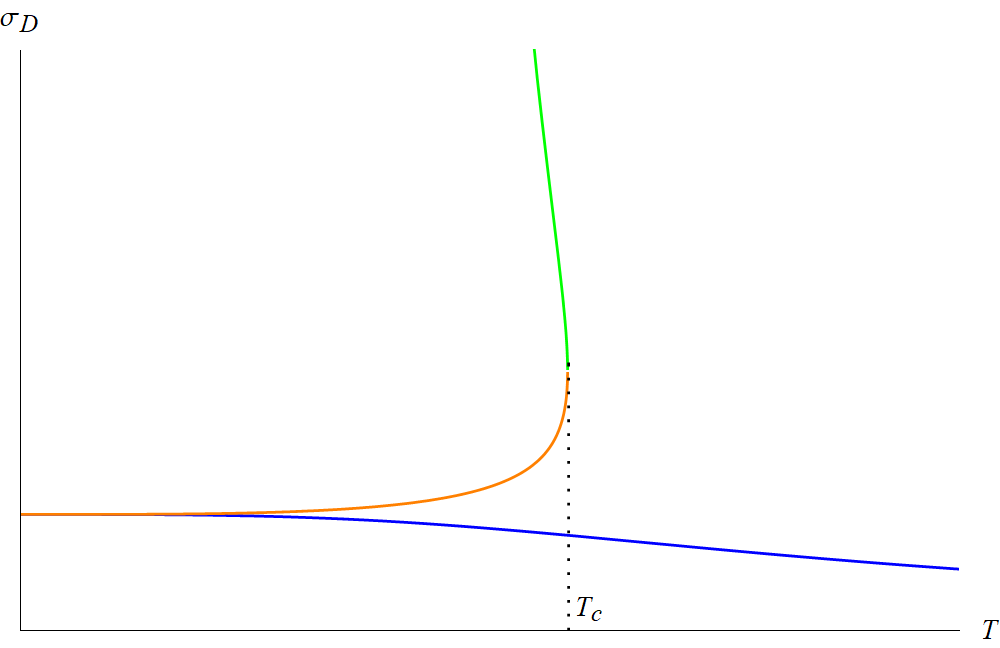}\label{fig:DCT3d4}}
\subfigure[{Plot of $\sigma_{D}\left(  T\right)  $ in the the blue, orange and green
phases for $d=12$. For $T>T_{c}$, jumping from one value of $\sigma_{D}$ to
another represents a first order phase transition.}]{
\includegraphics[width=0.45\textwidth]{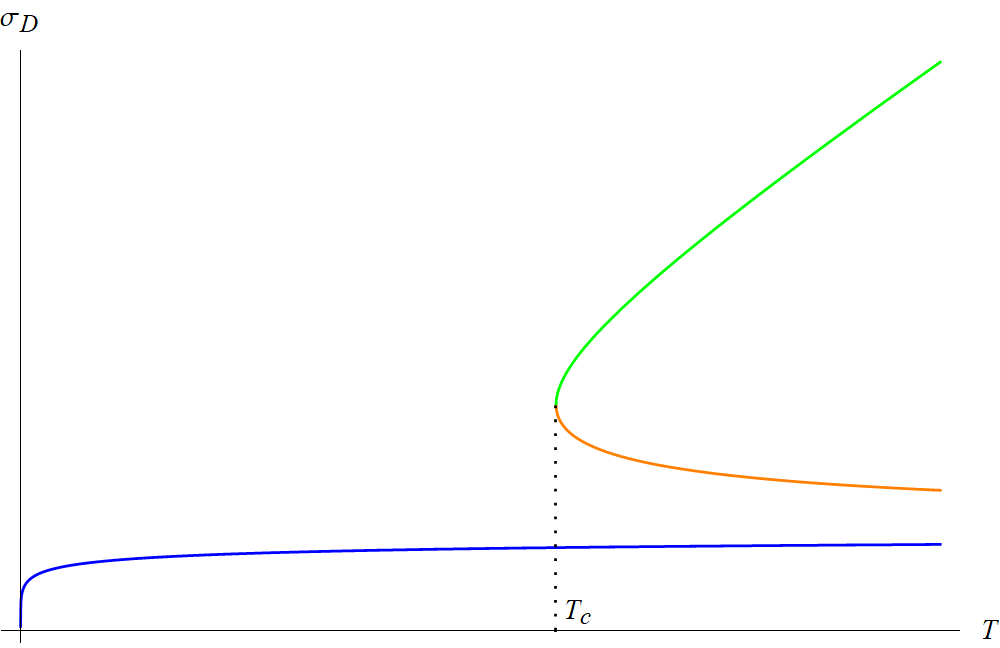}\label{fig:DCT3d12}}
\end{center}
\caption{Plot of $\sigma_{D}$ versus $T$ for $d=4$ and $12$ in the case of
$p=3$. Here the values of $\rho$ and $B$ are fixed with $\rho/B^{3}>0$, for
which the blue, orange and green phases could exist.}%
\label{fig:DCT3}%
\end{figure}

\begin{table}[h]
\begin{center}
{\scriptsize $%
\begin{tabular}
[c]{|c|c|c|c|}\hline
& Even $p$ & Purple, Green and Blue Phases for Odd $p$ & Orange and Red Phases
for Odd $p$\\\hline
$d<4p-1$ & Metal $\left(  \partial\sigma_{D}/\partial T<0\right)  $ & Metal
$\left(  \partial\sigma_{D}/\partial T<0\right)  $ & Insulator $\left(
\partial\sigma_{D}/\partial T>0\right)  $\\\hline
$d>4p-1$ & Insulator $\left(  \partial\sigma_{D}/\partial T>0\right)  $ &
Insulator $\left(  \partial\sigma_{D}/\partial T>0\right)  $ & Metal $\left(
\partial\sigma_{D}/\partial T<0\right)  $\\\hline
$d=4p-1$ & $\partial\sigma_{D}/\partial T=0$ & $\partial\sigma_{D}/\partial
T=0$ & $\partial\sigma_{D}/\partial T=0$\\\hline
\end{tabular}
\ \ \ \ $ }
\end{center}
\caption{Sign of $\partial\sigma_{D}/\partial T$ in all cases.}%
\label{tab:result}%
\end{table}

When $p$ is odd, one has for $y\ll1$ that $x\left(  y\right)  \sim1$ in the
purple and red phases; $x\left(  y\right)  \sim-1$ in the blue and orange
phases; $x\left(  y\right)  \sim-y$ in the green phase. In the case with
$\rho=0$ and nonzero $B$, we find that%
\begin{equation}
\sigma_{D}=\frac{pB^{2p-2}}{2^{p-1}}\left(  \frac{4\pi T}{3}\right)
^{d-4p+1}\text{ in the green phase, and }\sigma_{D}=0\text{ otherwise.}%
\end{equation}
From the monotonicity of $x\left(  y\right)  $, we can also determine whether
each phase is a metal or an insulator. The results are summarized in TABLE
\ref{tab:result}. In FIG. \ref{fig:DCT3}, we plot $\sigma_{D}$ versus $T$ for
$d=4$ and $12$ in the case of $p=3$. In FIG. \ref{fig:DCT3}, we fix the values
of $\rho$ and $B$ with $\rho/B^{3}>0$, for which only the blue, orange and
green phases exist. When $d=4$, FIG. \ref{fig:DCT3d4} shows that there are
three values for $\sigma_{D}$ for $T<T_{c}$, and jumping from one value to
another could represent a first order phase transition.\ Specially, if the
system jumps from the blue phase to the orange one or vice versa, one would
have a first order metal-insulator transition. A similar behavior applies to
$\sigma_{D}$ for $T>T_{c}$ in FIG. \ref{fig:DCT3d12}, where $d=12$.

\section{Discussion and Conclusion}

\label{Sec:Con}

In this paper, we extended the method of \cite{IN-Iqbal:2008by} to study the
electrical transport behavior of some boundary field theory in the presence of
a power-law Maxwell gauge field. In particular, we first calculated the
conductivities of the stretched horizon of some general static and neutral
black brane in the framework of the membrane paradigm. Since the conjugate
momentum of the power-law Maxwell field encoded the information about the
conductivities both on the stretched horizon and in the boundary theory and,
in the zero momentum limit, did not evolve in the radial direction, we
obtained the DC conductivity of the dual conserved current in the boundary
theory. We also found that the DC conductivity could be expressed in terms of
the electromagnetic quantities and the temperature of the boundary theory.

In the context of the membrane paradigm, we found that the second law of
black-hole mechanics required that the DC conductivities of the stretched
horizon and in the boundary theory are real and non-negative. Imposing
$\sigma_{D}\geq0$, we showed that, when $p$ was an even integer, there might
be two phases in the boundary theory, and a continuous phase transition could
occur at $\tilde{\rho}=0$. When $p$ was an odd integer, there might be five
phases in the boundary theory, and the transitions among them could be
considered as first order phase transitions. Specifically, it showed that the
green phase possessed a charge conjugation symmetric contribution, a negative
magneto-resistance and Mott-like behavior. We also discussed the temperature
dependence of the DC conductivity. We found that the DC conductivity
$\sigma_{D}$ was independent of the temperature of the boundary theory when
$d=4p-1$. Note that the power-law Maxwell field action is conformally
invariant for $d=4p-1$.

Finally, we discuss the assumption and limitation of our calculations. First,
we assumed that the black brane background was neutral, and hence there was no
background charge density in the boundary theory. Since the low frequency
behavior of the conductivities depends crucially on whether there is a
background charge density \cite{IN-Blake:2013bqa}, investigating the behavior
of the DC conductivity in a boundary theory dual to a charged power-law
Maxwell field black hole is certainly interesting. Second, we assumed that the
power-law Maxwell field was a probe field and neglected the backreaction on
the bulk spacetime metric. One would like to study the effects of backreaction
on the bulk spacetime metric and DC conductivity in the boundary theory.
Third, we carried out our calculations in the zero momentum limit, in which
the conjugate momentum did not evolve along the radial direction in bulk, and
the electromagnetic quantities $\rho$ and $B$ were time independent and
homogeneous in the boundary theory.

\begin{acknowledgments}
We are grateful to Houwen Wu and Zheng Sun for useful discussions. This work
is supported in part by NSFC (Grant No. 11005016, 11175039 and 11375121).
\end{acknowledgments}

\appendix

\end{document}